\documentclass[
    ,final            
  ]
  {aipproc}

\layoutstyle{6x9}
\usepackage{epsfig}

\begin{document}

\title{On the Leading Order Hadronic Contribution to $(g-2)_\mu$}

\author{Kim Maltman}{
  address={Math \& Stats, York Univ., Toronto, CANADA, and
CSSM, Univ. of Adelaide, Adelaide, AUSTRALIA}}

\pacs{13.35.Bv,13.40.Gp,13.66.Bc,13.35.Dx}

\keywords{muon anomalous magnetic moment, $\tau$ decay, QCD sum rules}
%

\begin{abstract}
Sum rule constraints dominated by the independent
high-scale input, $\alpha_s(M_Z)$, are shown to be 
satisfied by $I=1$ spectral data from hadronic $\tau$ decays, but violated by
the pre-2005 electroproduction (EM) cross-section data. 
Determinations of the Standard Model (SM)
hadronic contribution to $(g-2)_\mu$ 
incorporating $\tau$ decay data are thus favored
over those based solely on EM data, implying
a SM prediction for $(g-2)_\mu$ in agreement
with current experimental results.
\end{abstract}

\maketitle


After accounting for the accurately known,
purely leptonic contributions, the largest of the SM contributions
to $a_\mu \equiv (g-2)_\mu /2$ is that due to leading (LO)
hadronic vacuum polarization (VP), $\left[ a_\mu\right]_{had}^{LO}$.
This contribution can be expressed as a weighted integral, with known kernel,
over the electromagnetic (EM) spectral function $\rho_{EM}(s)$.
The uncertainty on $\left[ a_\mu\right]_{had}^{LO}$ which results
is at the $\sim 1\%$ level (comparable to the $0.5$ ppm 
experimental uncertainty on $a_\mu$~\cite{bnl})
and dominates the uncertainty in the SM prediction for $a_\mu$~\cite{review}.
This determination can, in principle, be improved using CVC, together
with data on the $I=1$ spectral function, $\rho_{I=1}(s)$, 
obtained from hadronic $\tau$ decay. (Isospin-breaking (IB) 
corrections~\cite{cen,kmcw} are needed for accuracy at the $1\%$ level.) 
Unfortunately, the IB-corrected $\tau$ data is in significant disagreement with
the corresponding $I=1$ EM data, 
$\rho^{EM}_{I=1}$ lying uniformly below $\rho_{I=1}^\tau$ 
in both the $4\pi$ region
and that part of the $2\pi$ region 
above the $\rho$ but below $1\ {\rm GeV}^2$~\cite{dehz}.
The determination of $\left[ a_\mu\right]_{had}^{LO}$
employing only EM data leads to a SM $a_\mu$ prediction
$\sim 2.5\sigma$ below experiment, while the alternate
determination incorporating $\tau$ decay data is compatible
with experiment at the $\sim 1\sigma$ level~\cite{dehz}.
We show here that sum rule constraints strongly
favor the $\tau$-based determination.

For $w(s)$ any function analytic in the region $\vert s\vert <M$, with $M>s_0$,
the $I=1$ vector (V) current and EM correlators, $\Pi (s)$, and
corresponding spectral functions, $\rho (s)$,
satisfy the FESR relations
$\int_0^{s_0}w(s)\, \rho(s)\, ds\, =\, -{\frac{1}{2\pi i}}\oint_{\vert
s\vert =s_0}w(s)\, \Pi (s)\, ds$.
To suppress duality violation and allow the use of the OPE 
on the RHS, $w(s)$ should satisfy $w(s=s_0)=0$~\cite{kmfesr}. At scales of
$\sim$ a few GeV$^2$, the OPE representation for V current
correlators is essentially
entirely dominated by its leading $D=0$ component, and hence
determined by the single input parameter $\alpha_s(M_Z)$, whose value is
known from independent high-scale studies~\cite{pdg04rev}.
If one works with weights $w(y)$, $y=s/s_0$, which satisfy
$w(1)=0$, and are non-negative and monotonically decreasing for
$0<y<1$, the fact that the EM version of $\rho_{I=1}$ lies uniformly
below the corresponding $\tau$ version implies that
\begin{itemize}
\item the normalization {\it and} slope with respect to $s_0$ 
of the EM-based spectral integrals, for a given $w(y)$, should be too
low relative to OPE predictions if it is the $\tau$-based data
which is correct and 
\item the normalization {\it and} slope with respect to $s_0$ 
of the $\tau$-based spectral integrals for a given $w(y)$ should be 
similarly too high if it is the EM-based data which is correct.
\end{itemize}

Sum rule tests of the two data sets have been 
performed for a number of different $w(y)$. See Ref.~\cite{kmgm2} for details 
on the OPE and spectral integral inputs (including short- and 
long-distance IB corrections for the $\tau$ data 
and input relevant to the small $D>0$ OPE contributions). 
It is found that both the normalization and slope with respect to
$s_0$ of the $\tau$ based spectral integrals are in excellent
agreement with OPE expectations, while both the normalization
and slope of the EM-based spectral integrals are low, particularly
if one uses the pre-2005 EM $\pi\pi$ spectral data.

Table~\ref{table1} quantifies the EM normalization problem, giving
the values of $\alpha_s(M_Z)$ needed to bring the OPE and spectral 
integrals into agreement for the $\tau$ and EM cases, at the
maximum common available scale, $s_0=m_\tau^2$. Results are shown for
$w(y)=w_1(y)=1-y$ and $w_6(y)=1-6y/5+y^6/5$ (which
have zeros of order $1,\, 2$, respectively, at $s=s_0$). The entries are
to be compared, e.g., to the high-scale average,
$\alpha_s(M_Z)=0.1195\pm 0.0016$, obtained by
excluding (i) $\tau$ decay input and (ii) the erroneous
heavy quarkonium input~\cite{qwg04} from the PDG04 average.
Table~\ref{table2} similarly quantifies the EM slope problem.
In the OPE column, ``indep'' and ``fit'' label results obtained using
(i) the independent high-scale $\alpha_s(M_Z)$ input and (ii) the 
fitted $\alpha_s(M_Z)$ values from Table~\ref{table1}, respectively.
We see that lowering the input $\alpha_s(M_Z)$ value to accommodate
the EM spectral integral normalizations at $s_0\sim m_\tau^2$ 
does {\it not} resolve the slope problem for the EM spectral integrals.

\begin{figure}
\epsfig{figure=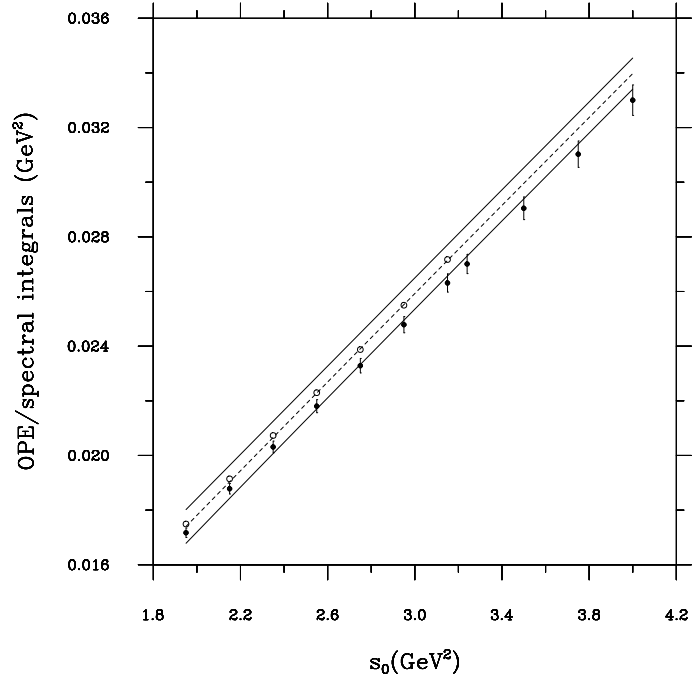, height=10cm,width=10cm}
  \caption{\label{fig1}OPE vs. spectral integrals for the $w_6$ FESR. All
notation as described in the text.}
\end{figure}

For illustration purposes,
the results for the $w_6$ EM case are also displayed in Fig.~\ref{fig1}.
The dotted and solid lines give
the central OPE and OPE error bounds, respectively, the solid
dots (with error bars) the EM spectral integrals, with
pre-2005 $\pi\pi$ input. The open circles show the shifted
``EM'' spectral integrals obtained by replacing the EM $2\pi$ and $4\pi$ 
data with the equivalent $\tau$ data. 
We see that both the EM slope and normalization
problems are resolved if, where the $I=1$ V part of the EM and
$\tau$ data disagree, the $\tau$ data are taken to be correct.
The recent 2005 SND EM $\pi\pi$ results~\cite{snd05} are compatible with the
corresponding $\tau$ results, further strengthening the case for the 
reliability of the $\tau$ data, and the agreement between the
SM prediction and experimental result for $a_\mu$.


\begin{table}
\caption{\label{table1}$\alpha_s(M_Z)$ from
fits to the $s_0=m_\tau^2$ experimental EM and $\tau$ spectral integrals, 
with central $D=2,4$ OPE input values}

\begin{tabular}{ccc}
\hline
Weight&\qquad\qquad$\left[ \alpha_s(M_Z)\right]_{EM}$\qquad\qquad&
\qquad\qquad$\left[ \alpha_s(M_Z)\right]_{\tau}$\qquad\qquad\\
\hline
$w_1$&$0.1138^{+0.0030}_{-0.0035}$&$0.1212^{+0.0027}_{-0.0032}$\\
$w_6$&$0.1150^{+0.0022}_{-0.0026}$&$0.1195^{+0.0020}_{-0.0022}$\\
\hline
\end{tabular}
\end{table}

\begin{table}[ht]
\caption{\label{table2}Slopes wrt $s_0$ of the EM OPE and spectral integrals}

{\begin{tabular}{cccc}
\hline
Weight&$S_{exp}$&$\alpha_s(M_Z)$&$S_{OPE}$\\
\hline
$w_1$&\ \ $.00872\pm .00026$\ \ &indep&\ \ $.00943\pm .00008$\ \ \\
&&fit&\ \ $.00934\pm .00008$\ \ \\
\hline
$w_6$&\ \ $.00762\pm .00017$\ \ &indep&\ \ $.00811\pm .00009$\ \ \\
&&fit&\ \ $.00805\pm .00009$\ \ \\
\hline
\end{tabular}}
\end{table}

\begin{theacknowledgments}
Thanks to the Natural Sciences and Engineering Research
Council of Canada and those authors
listed in the acknowledgements of Ref.~\cite{kmgm2}
for their input/assistance.
\end{theacknowledgments}


\bibliographystyle{aipproc}   


\IfFileExists{\jobname.bbl}{}
 {\typeout{}
  \typeout{******************************************}
  \typeout{** Please run "bibtex \jobname" to optain}
  \typeout{** the bibliography and then re-run LaTeX}
  \typeout{** twice to fix the references!}
  \typeout{******************************************}
  \typeout{}
 }

\end{document}